# Effect of $n^+$-GaAs thickness and doping density on spin injection of GaMnAs/$n^+$-GaAs Esaki tunnel junction

M. Kohda[a], Y. Ohno[a,b], F. Matsukura[a,c], and H. Ohno[a,c]

[a]*Laboratory for Nanoelectronics and Spintronics, Research Institute of Electrical Communication, Tohoku University, Japan*

[b]*CREST, Japan Science and Technology Agency, Japan*

[c] *ERATO Semiconductor Spintronics Project, Japan Science and Technology Agency, Japan*

**Abstract**

We investigated the influence of $n^+$-GaAs thickness and doping density of GaMnAs/$n^+$-GaAs Esaki tunnel junction on the efficiency of the electrical electron spin injection. We prepared seven samples of GaMnAs/$n^+$-GaAs tunnel junctions with different $n^+$-GaAs thickness and doping density grown on identical p-AlGaAs/p-GaAs/n-AlGaAs light emitting diode (LED) structures. Electroluminescence (EL) polarization of the surface emission was measured under the Faraday configuration with external magnetic field. All samples have the bias dependence of the EL polarization, and higher EL polarization is obtained in samples in which $n^+$-GaAs is completely depleted at zero bias. The EL polarization is found to be sensitive to the bias condition for both the (Ga,Mn)As/$n^+$-GaAs tunnel junction and the LED structure.

*PACS:* 75.25.Dc, 72.25.Hg, 73.40.Gk, 75.50.Pp

Keywords: (Ga,Mn)As, spin injection, Esaki diode, ferromagnetic semiconductor

## 1. Introduction

Efficient electrical injection of spin polarized electrons is one of the important technologies for realizing new functional devices based on spins in semiconductors [1]. A ferromagnetic semiconductor (Ga,Mn)As is one of the promising materials for spin-polarized carrier injector because of its high spin polarization [2] as well as high quality heterojunction with GaAs, but it is p-type with high hole concentration. Thus, electrical *electron* spin injection into non-magnetic semiconductors cannot be achieved simply by making a heterojunction using (Ga,Mn)As and applying an electric field. Instead of diffusive injection of spin polarized holes from (Ga,Mn)As to adjacent GaAs [3], interband tunnel injection of spin polarized electrons from the valence band in (Ga,Mn)As into the conduction band of n-GaAs has been demonstrated by using a GaMnAs/$n^+$-GaAs Esaki tunnel junction [4-6]. The spin polarized electron current is detected by measuring the electro-luminescence (EL) polarization from the embedded light emitting diode (LED), where the spin polarized electrons recombine with unpolarized holes.

Based on this scheme, the highest EL polarization of ~20%, which corresponds to ~80% of polarization of injected electron spins, has been reported so far [6], suggesting that this approach is quite useful for high efficiency spin injection. It has also been reported, however, that the spin injection efficiency is sensitive to the bias voltage [6] applied forward to the LED structures but reversely to the (Ga,Mn)As/$n^+$-GaAs junction in series, which is not the case in our earlier report [4]. This suggests that the mechanism of the interband tunnel spin injection strongly depends on the structural parameters, and thus the control of the electronic states at the interface should be critical for achieving efficient spin injection.

In the present work, we investigated the dependence of the EL polarization and the spin injection efficiency on the thickness and the doping density of $n^+$-GaAs at GaMnAs/$n^+$-GaAs tunnel junction.

## 2. Sample preparation

For the present study, we prepared samples with identical structures except for the thickness and the doping density of $n^+$-GaAs at the tunnel junction. Figure 1 illustrates the sample structures and the device geometry. All the samples were grown on p-GaAs (001) substrates by molecular beam epitaxy with overgrowth technique [3,4]. The Esaki tunnel junction consists of a 20 nm $(Ga_{0.943}Mn_{0.057})As$ and a $d$ nm of $n^+$-GaAs with Si doping density of $N_d$ $cm^{-3}$.

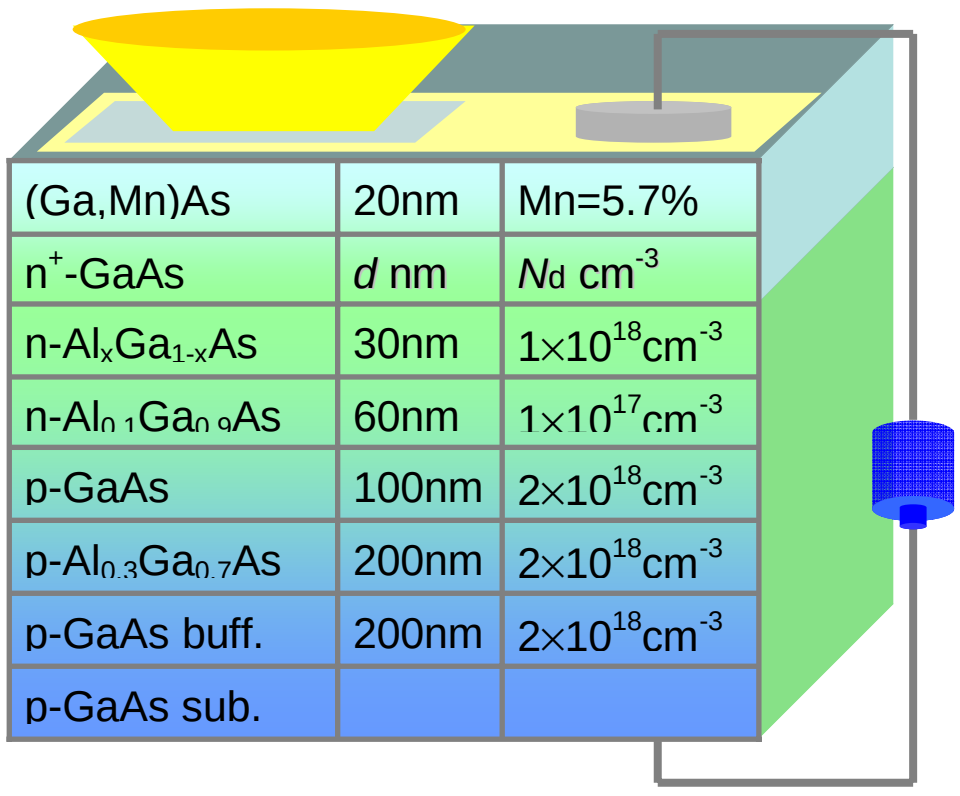

Fig. 1 The cross section of the sample structures and the device geometry is schematically shown. The thickness $d$ [nm] and Si doping density $N_d$ [$cm^{-3}$] of $n^+$-GaAs are listed in Table I.

Table I The structural parameters (the thickness of $n^+$-GaAs $d$ [nm] and the doping density $N_d$ [$cm^{-3}$]), and $V_{th}$ [V] are listed.

| Sample ID | A | B | C | D | E | F | G |
|---|---|---|---|---|---|---|---|
| $d$ [nm] | 8 | 10 | 15 | 10 | 15 | 20 | 15 |
| $N_d$ [$\times10^{19}cm^{-3}$] | 1.5 | 1.5 | 1.5 | 1.0 | 1.0 | 1.0 | 0.6 |
| $V_{th}$ [V] | 2.2 | 1.5 | 1.5 | 2.0 | 1.5 | 1.5 | 2.5 |

The Curie temperature of (Ga,Mn)As of the present sample is 70 K. The tunnel junction was formed on the top of a spin-detection LED similar to that in Ref. 6: it consists of a 30 nm graded n-$Al_xGa_{1-x}As$ ($x$ = 0~0.1) with the Si density of $1\times10^{18}$ $cm^{-3}$, a 60 nm n-$Al_{0.1}Ga_{0.9}As$ spacer ($1\times10^{17}$ $cm^{-3}$), a 100 nm p-GaAs acitive layer ($2\times10^{18}$) $cm^{-3}$, and p-$Al_{0.3}Ga_{0.7}As$/p-GaAs buffer layers. Seven samples of GaMnAs/$n^+$-GaAs tunnel junctions with different $n^+$-GaAs thickness ($d$ = 8, 10, 15 and 20 nm) and doping density ($N_d$ = 0.6, 1.0 and $1.5\times10^{19}cm^{-3}$) prepared are labeled A-G as listed in the Table I.

The epilayers were processed to 600 μm×600 μm mesas. The top Au/Cr electrode with a window for light emission was formed by a standard photolithograph and lift-off techniques. Bias voltage can be applied between the top electrode and $p^+$-GaAs substrate, and the light emitted from the top window is collected.

## 3. *I-V* characteristics

We first studied the current-voltage (*I-V*) characteristics of these devices at 10 K without applying magnetic field. In Fig. 2, the *I-V* curves of the samples A-C with $N_d = 1.5\times10^{19}$ $cm^{-3}$ are plotted in a semi-log scale. It is clearly seen that all these curves show kink, indicated by ▲, as the bias voltage is increased: Hereafter we refer to the kink point as $V_{th}$. It is clearly seen that $V_{th}$ is shifted to the higher bias as $d$ becomes smaller. We also measured *I-V* characteristics of samples D-G: All the *I-V* curves have similar features as shown in Fig. 2. For all the samples, $V_{th}$ are shown in Table I.

Assuming the hole concentration of (Ga,Mn)As to be $10^{20}$ $cm^{-3}$ [7], we solved the Poisson equation and calculated the depletion lengths in $n^+$-GaAs for three different doping densities: They are 11, 14, and 19 nm for $1.5\times10^{19}$, $1.0\times10^{19}$, and $6\times10^{18}$ cm$^{-3}$, respectively. As seen in Table I, we found that samples C, E, and F, in which $d$ is thicker than the depletion length, have the same $V_{th}$ (= 1.5 V). On the other hand, $n^+$-GaAs should be fully depleted in samples A, D, and G with higher $V_{th}$ (> 2.0 V). An exception is the sample B, for which $V_{th}$ = 1.5 V but $d$ = 10 nm is thinner than the calculated depletion length (11 nm).

From this trend, it is obvious that the conductivity at the (Ga,Mn)As/$n^+$-GaAs tunnel junction is sufficiently high when a neutral $n^+$-GaAs is retained: $V_{bias}$ is mainly applied to the bottom LED pn junction up to $V_{th}$, which corresponds to the flat band condition. Above $V_{th}$, $V_{bias}$ shall be divided to the forward bias on LED and the reverse bias to the tunnel junctions so that the current through the two junctions can be matched.

In samples A, D, and G, on the other hand, $n^+$-GaAs should be fully depleted. This results in the distribution of $V_{bias}$ for both junctions at any $V_{bias}$ and thus higher $V_{th}$ (>1.5 V).

## 4. EL-polarization

The EL polarization of the surface emission was measured under the Faraday configuration with the external magnetic field $B$ applied parallel to the growth direction. The EL spectra were taken by using a CCD camera with a 50 cm spectrometer. The right- ($\sigma^+$) and the left- ($\sigma^-$) circular polarization components of the EL intensity, $I^{\sigma+}$ and $I^{\sigma-}$, were resolved by a liquid crystal retarder and a linear polarizer. The EL polarization is defined as $(I^{\sigma+} - I^{\sigma-})/(I^{\sigma+} + I^{\sigma-})$.

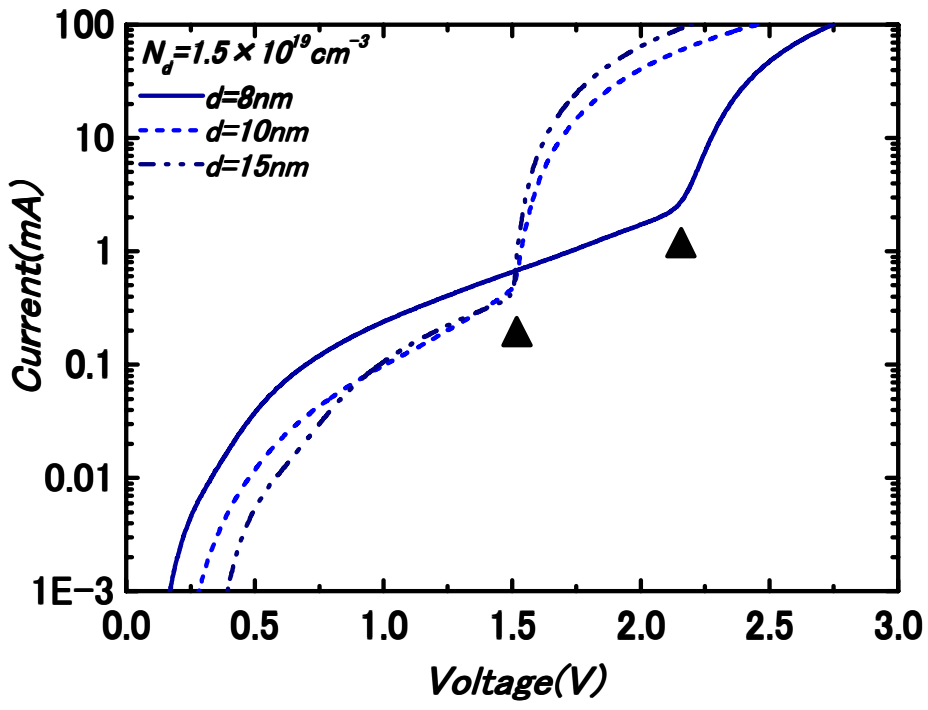

Fig. 2 The current-voltage characteristics of samples A ($d$ = 8 nm), B (10 nm), and C (15 nm) measured at 10 K. ▲ indicates the kink point (referred to as $V_{th}$ in the text).

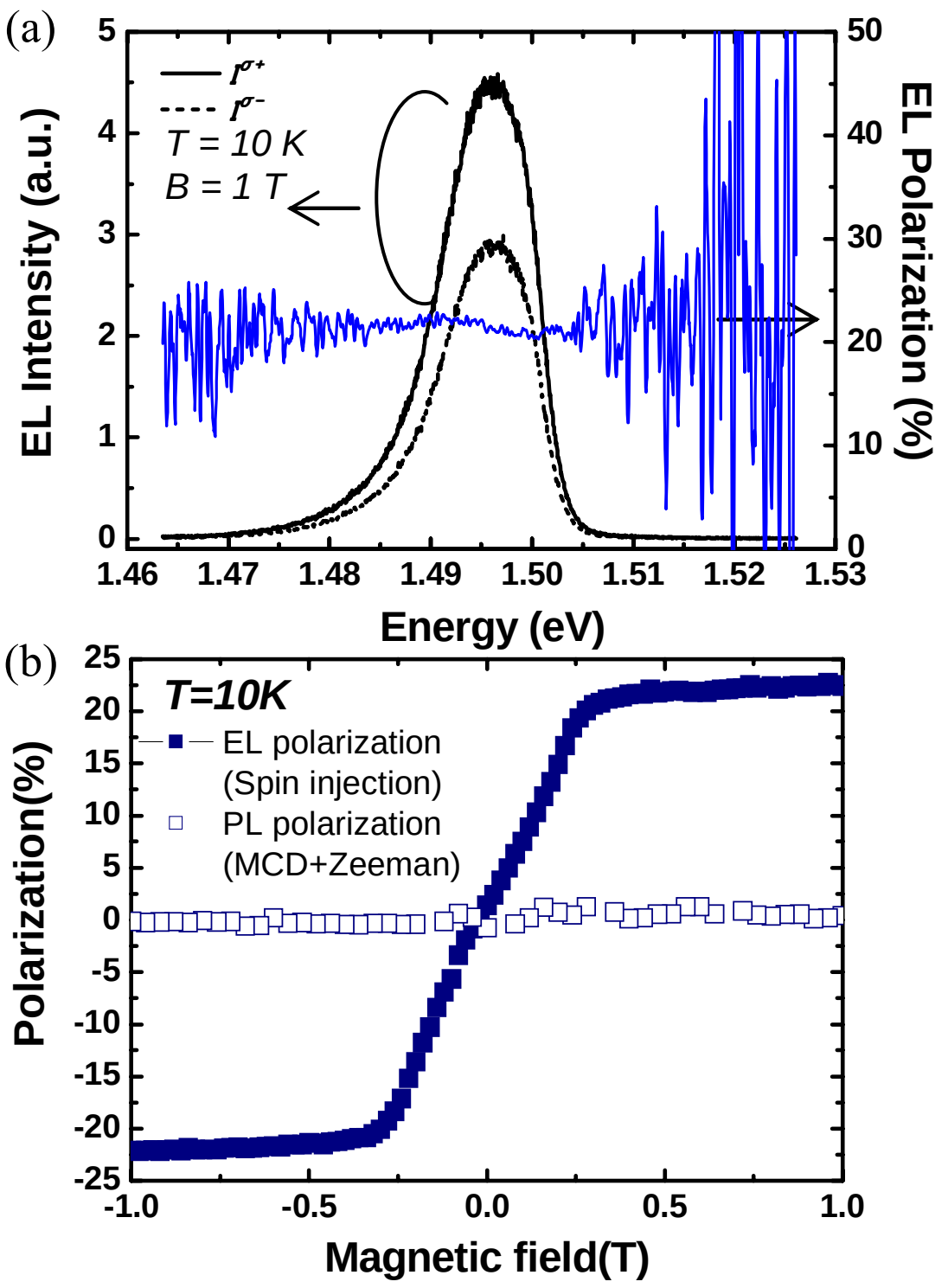

Fig. 3 (a) The EL spectra ($\sigma^+$ and $\sigma^-$ polarization) and EL polarization of sample D at $B$ = 1 T and $V_{bias}$ = 1.85 V. (b) Magnetic field dependence of EL polarization (closed symbols) measured at 10 K. The open symbols are the photo-luminescence (PL) polarization measured in the same sample excited by linear polarized light from the top.

Figure 3(a) shows the EL spectra and the EL polarization of sample D measured at 10 K, $B$ = 1 T, and $V_{bias}$ = 1.85 V. This gives the highest EL polarization (22.3%) in the present study. Closed symbols in Fig. 3(b) show $B$-dependence of the EL polarization taken under the same condition as that in Fig. 3(a). For $B$ > 1T, the EL polarization saturates as the magnetization of (Ga,Mn)As. The open symbols shown in Fig. 3(b) are the photoluminescence (PL) polarization, which was taken by exciting the sample with linear polarized light from the top and without applying $V_{bias}$. This gives the background polarization induced by the magnetic circular dichroism (MCD) of (Ga,Mn)As and the Zeeman spin splitting in the active layer. As shown in Fig. 3(b), it is found to be 0.2% at $B$ = 1 T. Thus the EL polarization due to the electrical spin injection is corrected to be 22.1% by subtracting 0.2% from the observed EL polarization.

Next, we examined the $V_{bias}$ dependence of the EL polarization. Figure 4 shows the EL polarization as a function of $V_{bias}$ for all the samples. We found different $V_{bias}$ dependence of the EL polarization: For samples A, D, and G, which have higher $V_{th}$, a distinct peak (>15%) was observed at 1.7< $V_{bias}$<1.9 V. Note that $V_{bias}$ at the peak of EL polarization is less than $V_{th}$, i.e. in the pn junction of LED there remains a potential barrier for injection of minority carriers. As $V_{bias}$ is increased further, the EL polarization is reduced to < 5%. For the other samples, on the other hand, the EL polarization gradually increases with $V_{bias}$ up to 4~8%, but then tends to saturate or decrease.

These results indicate that the efficiency of the spin injection is quite sensitive to $V_{bias}$ depending on the thickness and the doping density of $n^+$-GaAs layer: it is concluded from the experimental data that as long as the level of tunnel injection current is not high, higher degree of spin polarization is retained. When higher reverse bias voltage is applied to the (Ga,Mn)As/$n^+$-GaAs tunnel junction to feed larger current to the LED, on the other hand, unpolarized valence electrons existing deep in the valence bands of (Ga,Mn)As can tunnel into the adjacent GaAs conduction band, which results in degradation of the efficiency of the electrical *electron* spin injection.

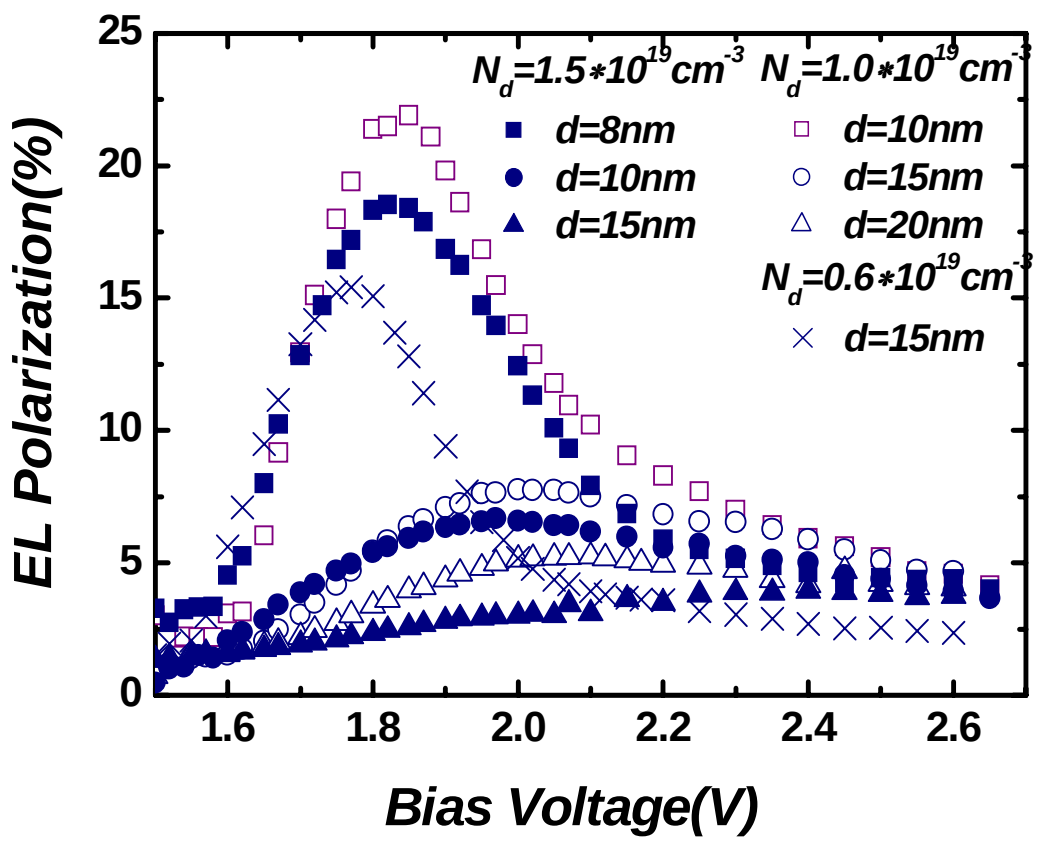

Fig. 4 Bias voltage dependence of the EL polarization measured at 10 K, 1 T.

## 5. Conclusion

In conclusion, we investigated the effect of $n^+$-GaAs thickness and doping density on the efficiency of electrical electron spin injection through GaMnAs/$n^+$-GaAs junctions via interband tunnelling. The maximum EL polarization of 22.1% was obtained. High EL polarization was observed in samples in which $n^+$-GaAs layer is completely depleted at zero bias voltage. The EL polarization is found to be sensitive to the bias condition for both the (Ga,Mn)As/$n^+$-GaAs tunnel junction and the LED structure.

This work was partly supported by Grant-in-Aids from JSPS and MEXT.